\documentclass[
reprint,
amsmath,
mssymb,
prper,
floatfix,
twocolumn, superscriptaddress,
]{revtex4-1}

\usepackage{array, makecell}
\usepackage{graphicx}
\usepackage{dcolumn}
\usepackage{bm}
\usepackage{scalerel,amssymb}
\def\msquare{\mathord{\scalerel*{\Box}{gX}}}

\begin{document}

\preprint{APS/123-QED}

\title{How Physics Textbooks Embed Meaning in the Equals Sign}
\author{Dina Zohrabi Alaee}
\affiliation{%
 School of Physics and Astronomy, Rochester Institute of Technology, Rochester, New York, USA.}%
\author{Eleanor C. Sayre*}
 \affiliation{%
 Department of Physics, Kansas State University, Manhattan, Kansas 66506, USA}%
\affiliation{%
 School of Physics and Astronomy, Rochester Institute of Technology, Rochester, New York, USA.}%
 \email{esayre@gmail.com}
\author{Kellianne Kornick}
\affiliation{%
 School of Physics and Astronomy, Rochester Institute of Technology, Rochester, New York, USA.}%
\author{Scott V. Franklin}
\affiliation{%
 School of Physics and Astronomy, Rochester Institute of Technology, Rochester, New York, USA.}%
\date{\today}

\begin{abstract}
Physics as a discipline embeds conceptual meaning about the physical world in mathematical formalism. The meaning associated with mathematical symbols depends on context, and physicists can shift conceptual meaning by manipulating those symbols. We present an analysis of the different physical meanings associated with the equal sign ``='' that can be inferred from introductory and upper--level physics textbooks. Five distinct meanings/categories are identified: causality, balancing, definitional, assignment, and calculation, each with operational definitions that help identify their presence. The different uses can be seen to link mathematical equations to intuitive conceptual ideas, and significant differences in the frequency with which these are used exist between textbooks of different levels.
\end{abstract}


\newcommand\myeqA{\mathrel{\stackrel{\makebox[0pt]{\mbox{\normalfont\tiny A}}}{=}}}
\newcommand\myeqM{\mathrel{\stackrel{\makebox[0pt]{\mbox{\normalfont\tiny M}}}{=}}}
\newcommand\myeqD{\mathrel{\stackrel{\makebox[0pt]{\mbox{\normalfont\tiny D}}}{=}}}
\newcommand\myeqB{\mathrel{\stackrel{\makebox[0pt]{\mbox{\normalfont\tiny B}}}{=}}}
\newcommand\myeqH{\mathrel{\stackrel{\makebox[0pt]{\mbox{\normalfont\tiny Hybrid}}}{=}}}
\newcommand\myeqC{\mathrel{\stackrel{\makebox[0pt]{\mbox{\normalfont\tiny C}}}{=}}}

\maketitle

\section{Introduction}
In recent years, the interest in mathematics as the language of physics has been growing. Taking up this metaphor, in this study we examine ``grammar'' on a minute level to investigate the particular dialect of mathematics (principally, the equal sign) as used in physics textbooks.\par
The concept of equality is surprisingly complex. Several studies have documented that students often misinterpret the equal sign as an operational, not relational, symbol~\cite{Knuth2006, Molina2007, Knuth2008, RittleJohnson2011, Sherman2009, Stephens2013}.

Understanding the equal sign in a relational manner is important due to its role in upper--level mathematics and physics courses, and so we seek a record of how equal signs are used across physics programs. For U.S. physics university curricula, this means focusing on textbooks. We look at five physics textbooks to investigate the language that authors as expert physicists use in a physics context. Focusing attention on the structure of the equations involving the equal sign leads to an understanding of an equation's underlying meaning which can then help illuminate the dialect of mathematics used in physics.

There is a long history of mathematics education research, mostly in K-12 contexts, into students' understanding of mathematical symbols in general and equality in particular~\cite{Renwick1932, Denmak1976, Behr1976, ginsburg1977, Behr1980, Kieran1981, BaroodyGinsburg1982, LudlowWalgamuth1998, Knuth2006, Oksuz2007, Knuth2008, Noonan2014, ByrdMcNeilChesneyMatthews2015}. In one of the earliest studies, Behr \textit{et al.}~\cite{Behr1976} observed that elementary school children
\textit{``consider the symbol ``='' as a ``do something signal'' that ``gives the answer'' on the right hand side. There is a strong tendency among all the children to view the ``='' symbol as being acceptable when one (or more) operation signs precede it.''}

Falkner \textit{et al.}~\cite{Falkner1999} identified kindergarten students that understood the concept of equality but could not transfer that understanding to algebraic problems. He also found that students often interpreted the equal sign as indicating action (a ``do it'' sign), with older students gradually recognizing it as a symbol that indicates a relationship. Knuth \textit{et al.}~\cite{Knuth2006} linked middle school students' understanding of the equal sign with performance on solving algebraic equations. These and other contemporaneous studies focus on the mathematical--appropriate abstractions of equality, using physical systems primarily as examples and illustrations. Other studies confirm that students across K-12 see the equal sign as primarily an operational symbol and do not have a deeper understanding of mathematical equivalence~\cite{Knuth2006, Kieran1981, BaroodyGinsburg1982, Knuth2006}. Kieran~\cite{Kieran1981} found that the idea of the equal sign as an operator is formed before formal education begins and continues throughout high school. This view encourages students to see formulas as knowledge to be memorized and prevents a recognition of the underlying meaning and structure.

Physics education research has documented student approaches to solving problems in specific physics contexts~\cite{Hellerp11991, Hellerp21991, Gabel1994, Thacker1994, Hsu2004, Meltzer2005}, examining how students form relevant representations to understand and communicate physical ideas to solve problems~\cite{Fredlund2014}. In order to translate a problem statement into algebraic expressions, students may encounter many different representations of physics ideas, including gestures~\cite{Scherr2008}, graphs and diagrams~\cite {Ottosson2006, Rosengrant2009, Fredlund2012, christensen2012}, mathematics~\cite{Sherin2001, Sherin2006, Ragout2002, Domert2007, bing2009}, and language~\cite{Brookes2006, Linder2006, Linder2011}. 

Most physics education research on problem solving has focused either on students' conceptual understanding or on engagement with mathematical processing~\cite{Thacker1994, Huffman1997,VanHeuvelen1991b, Heller1992a, Hsu2004, Meltzer2005, Walsh, Redish2008LookingEngineers, Reif2008ApplyingDomains,kh2013}, rarely connecting the two. In a review of over a decade of published articles on problem solving from nine leading physics and science education journals, Kuo \textit{et al.}~\cite{kh2013} found ``no studies that focused upon the mathematical processing step or described alternatives to using equations as computational tools.'' This is despite the general recognition that the interpretation of mathematical symbols is a necessary skill in developing students' understanding of physics~\cite{Leonard1996, Mazur1998, Redish2005}. Subsequently, Uhden \textit{et al.}~\cite{uhden2012} used the term ``mathematization'' in developing a model for how mathematics is used in physics education. A core feature of understanding students' mathematizing in physics is identifying how students represent concepts symbolically, verify solutions, and connect both to the physical world.\cite{Freudenthal1973,Adrian1987,brahmia2015,brahmia2016}. 

Sherin~\cite{Sherin2001,Sherin2006} proposed the \textit{symbolic form} as a cognitive mathematical primitive that associates physics conceptual meaning with mathematical symbols in order to understand ``how students understand physics equations.'' He observed that students associate various conceptual ideas with mathematical expressions as they solve problems and identify numerous different forms. We took up the idea of symbolic forms and focused on the conceptual meaning behind the equal sign. More broadly, we posit that the equal sign doesn't happen in isolation: the equal sign is an element of mathematical sentences. The forms of equations are context--dependent, and equivalent mathematical equations can have different symbolic forms. For example, the right hand side of the kinematic equation $v{_{f}} = v{_{0}} + at$ can be interpreted as a ``base$+$change'', with the initial velocity $v{_{0}}$ modified by the change in velocity brought about by acceleration. The topologically equivalent equation for net force of a spring-gravity system $F{_{net}} = -kx- mg$, however, is more likely to be interpreted as a ``sum of parts'', with the net force $F_{net}$ the sum of the various forces, in this case gravity $mg$ and spring $kx$. 

Tuminaro and Redish \textit{et al.}~\cite{Tuminaro2007} used symbolic forms to model how students translate mathematical solutions into physical understanding, with additional work from Kuo \textit{et al.}~\cite{kh2013} revealing that students do not expect conceptual knowledge of mathematics to connect to their problem solving. 

This study extends previous work~\cite{DZA2018} and explores the conceptual meaning behind mathematical formalisms. We analyze physics textbooks to investigate the disciplinary interpretation of the equal sign ``=''. In doing so, we do not ask how the ``='' understanding might be used to solve a problem, but rather whether thematic categories arise that are plausible to a physicist's interpretation of the symbol. Our method parallels that of Burton \textit{et al.}~\cite{Burton2000} who studied published journal articles in a variety of mathematical sub-fields to identify a ``natural language'' in their epistemological practice. We find a shared focus in the work of Kress~\cite{Kress} (in Cope and Kalantzis book) in striving to understand ``what language [including, in our case, math symbols] is doing and being made to do by people in specific situations in order to make particular meanings'' and agree with Burton \textit{et al.}~\cite{Burton2000} that doing so may ``shed some light on the values and meanings of the practices'' of physicists in the pedagogical context.

\section{Textbook selection}

Our study focuses on five textbooks (Table~\ref{tab:1}) spanning introductory through senior--level coursework in Mechanics, Electrostatics, and Quantum Mechanics. Physics curricula are often cyclical, and later courses often return to previously covered material with more depth and mathematical sophistication. Because of this, we selected chapters with similar content, allowing us to see differences across both topic and level. 

At the introductory level, we analyze \textit{University Physics with Modern Physics} (14\textsuperscript{th} edition)~\cite{Young2015}, a popular introductory physics textbook used at universities around the world. Chapters 21 and 22 of this text focus on electric charge, electric field, and Gauss' law. 

At the middle division, we analyze \textit{Modern Physics}~\cite{Krane1995} and \textit{Classical Mechanics}~\cite{Taylor2005}. \textit{Modern Physics} balances the concepts of quantum physics with their historical development as well as the experimental evidence supporting theory. Chapter 5 focuses on the wave behavior of particles, the time--independent Schr\"odinger equation, the ``particle in a box'' problem in one--dimension and two--dimensions, and the quantum harmonic oscillator. 
\textit{Classical Mechanics}~\cite{Taylor2005} covers Newton's laws of motion, projectiles and charged particles, momentum and angular momentum, energy, oscillations, and Lagrange's equations. Chapter 4 covers conservation of energy, central forces systems, energy of a multi--particle system, and elastic collisions.

At the upper division, we analyze \textit{Introduction to Electrodynamics}~\cite{Griffiths1999EM} and \textit{Introduction to Quantum Mechanics}~\cite{Griffiths2005QM}. These are the two most popular textbooks for their respective courses. \textit{Introduction to Electrodynamics} presents a strongly theoretical treatment of electricity and magnetism. Chapter 2 focuses on electrostatics and electric fields, particularly Coulomb's Law and Gauss' Law. \textit{Introduction to Quantum Mechanics} balances discussions of quantum theory with mathematical treatments from a wave functions-first perspective. Chapter 2 the time--independent Schr\"odinger equation for both the particle in a box and the harmonic oscillator.

Physics undergraduate textbooks in general are extremely consistent in content and presentation, suggesting that our results should be generalizable to other physics textbooks. 

\begin{table*}[t]
  \begin{center}
    \caption{Textbook selection}
    \label{tab:1}
    \begin{tabular}{l|l|c|p{9cm}} 
    \hline\hline
      \textbf{Textbook level} & \textbf{Textbook and authors} & \textbf{Chapters} & \textbf{Description}\\
      \hline
      Introductory & 
       \makecell[l]{University Physics with \\ Modern Physics, 14\textsuperscript{th} ed.\\ Young \& Freedman} & 21, 22 & \makecell[l]{Charge, Gauss's law, and electric field}\\
      \hline
      Intermediate & \makecell[l]{Modern Physics, 2\textsuperscript{nd} ed.\\ Kenneth S. Krane} & 5 & \makecell[l]{Time--independent Schr\"odinger equation and harmonic oscillator}\\
      & \makecell[l]{Classical Mechanics, 2005, \\John R. Taylor} & 4 & \makecell[l]{Central--force problems, non--inertial frames, coupled oscillators,\\ and nonlinear mechanics}\\
      \hline
      Upper--division & \makecell[l]{Introduction to \\ Electrodynamics, 4\textsuperscript{th} ed.\\ David J. Griffiths}  & 2 & \makecell[l]{Vector analysis, electrostatics, electric and magnetic fields,\\ electrodynamics, Coulomb's Law, and Gauss' Law}\\
      
      & \makecell[l]{Introduction to Quantum \\ Mechanics, 2\textsuperscript{nd} ed.\\ David J. Griffiths} & 2 & \makecell[l]{Wave functions, time--independent Schr\"odinger equation,\\ particle in a box, and the harmonic oscillator}\\
      \hline
    \end{tabular}
  \end{center}
\end{table*}

\section{Methodology}


The categorization scheme was developed through iterative readings of the textbooks. After reading each chapter, two researchers individually wrote down the key points they noticed about the equations. After working through three chapters, we created sets of notes that described each equation observed in a symbolic template, noting the conceptual meaning associated with the equal sign in that equation. The first draft of categories came from this data.

The researchers then carefully re-read each selected chapter to identify the category for each equal sign. After coding all the equations individually, results were discussed in a group to refine the articulations. Equations with similar meanings were grouped in order to develop robust descriptions of each of category. After several iterative cycles of analysis and refinement,  the coding scheme was judged stable and an outside researcher used the coding scheme on random sections from each chapter to establish inter rater reliability (IRR). \text{87.5\%} of initial coding overlapped with the original researchers. After clarifying discussions, including a tutorial about the code book and discussions with each equation, subsequent IRR tests resulted in \text{100\%} agreement.

\begin{figure}[!htp]

\raggedright{\noindent\bf{Example 2.3} \\ A long cylinder carries a charge density that is proportional to the distance from the axis: $\rho\myeqA ks$, for some constant $k$. Find the electric field inside this cylinder.}
\\[0.5\baselineskip]

\includegraphics[width=2.5in]{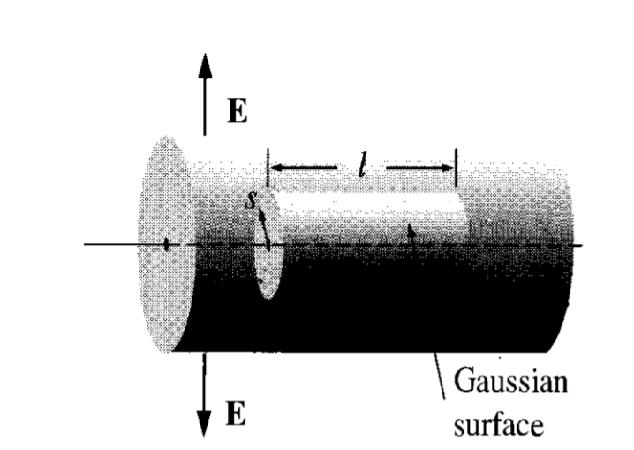}
\centering

\raggedright{\bf{Solution:} Draw a Gaussian cylinder of length $l$ and radius $s$. For this surface, Gauss's law states:}\\

\begin{eqnarray*}
\oint E.da \myeqD \frac{1}{\epsilon_0}Q_{enc}, \nonumber
\end{eqnarray*}

\raggedright{\bf The enclosed charge is}\\

\begin{eqnarray*}
Q_{enc} & \myeqD \int (\rho d\tau) \myeqA \int (ks')(s'ds' d\Phi dz)\myeqM 2\pi kl\int_{0}^{s} s'2ds'\\
& \myeqM \frac{2}{3} \pi kl s^{3}.
\end{eqnarray*}

Now, symmetry dictates that $E$ must point radially outward, so for the curved portion of the Gaussian cylinder we have:\\ 

\begin{eqnarray*}
\int Eda \myeqA \int \left |E \right | da \myeqM \left |E \right | \int da\myeqM \left | E \right |2\pi sl,
\end{eqnarray*}

While the two ends contribute nothing (here $E$ is perpendicular to $da$). Thus,
\begin{eqnarray*}
\left | E \right |2\pi sl \myeqA \frac{1}{\epsilon_0}\frac{2}{3}\pi kls^{3}
\end{eqnarray*}\\
Or finally,
\begin{eqnarray*}
E\myeqM \frac{1}{3\epsilon _0}ks^{2}\hat{s}
\end{eqnarray*}
\caption{Visual depiction of coding of Example 2.3 from \textit{Introduction to Electrodynamics}}
\label{Fig1}
\end{figure}

An example of the coding applied to a problem in the textbook is shown in Figure \ref{Fig1}, which shows a problem as stated in the textbook with a worked solution, including both equations and descriptive text. Every equal sign is assigned a code that indicates its categorization ($\myeqD$ for Definition, $\myeqC$ for Causal, $\myeqA$ for Assignment, $\myeqB$ for Balancing, or $\myeqM$ for Calculation). We reiterate that, in this study, \textit{every} equal sign that appears in the selected chapter is assigned a unique code.

\section{Categories}
Five categories emerged from our study: definitional, causal, assignment, balancing and calculation. Table~\ref{tab:2} summarizes the five categories, including operational articulation, canonical form and direction.

\subsection{Definitional (D)}
As with most disciplines, physics uses careful definitions to constrain ideas to narrow and specific uses. The equal sign mediates this definition in mathematical expressions through an operational articulation ``is always''. For example, the equation (here and henceforth we omit vector signs for simplicity)
\begin{equation} \label{eq1}
m=\frac{F{_{net}}}{a}
\end{equation}  defines the inertial mass $m$ as the ratio of net force to resulting acceleration. This definition is always true in the context of mechanics. A variation of the definition is used to define a mathematical formalism:
\begin{equation} \label{eq2}
\lim_{\Delta t\to0 }\frac{\Delta v_{x}}{\Delta t}=\frac{dv_{x}}{dt}
\end{equation} 

The order in which an equation is read is important. Rittle--Johnson~\cite{rittle1998} has found that elementary-school children read all equations left-to-right, whereas physicists read in specific directions depending on their contextual use. Definitional equations are read left-to-right: ``inertial mass is defined as the ratio of net force to acceleration'' and ``the derivative is defined as the limit...''

\subsection{Causality (C)}
Much of physics involves inferring causal relationships. Forces cause (operationally ``lead to'' or ``result in'') accelerations, and charged particles or currents cause electric or magnetic fields respectively. Examples of equations that indicate causal relationships include 
\begin{equation} 
\label{eq3}
a=\frac{F{_{net}}}{m}
\end{equation} (forces cause accelerations)

In our discussions within the research team and with community members, the distinctions between \textit{causal} and \textit{definitional} equal signs were  dependent on context, moreso than for any other two categories.  Sometimes, multiple codes were assigned to an equal sign, depending on the other parts of the solution and context.  

To distinguish these two, we turn to mechanistic explanations~\citep{Russ2012}.   For example, to describe the electric field, $E$, we build a mechanistic story to that equation such as: Electric fields are created by electric charges. The charges exert a force on one another by means of the disturbances that they generate in the space surrounding them. These disruptions are called electric fields. The electric field generated by a set of charges can be measured by putting a point charge $q$ at a given position.  In math, we might express this story with a causal equal sign.  

In contrast, a definitional equal sign does not need for such a description each time the equation is used~\citep{Sayre2008Coords}. We note that the causal agents are customarily placed on the right side of the equation and the resulting quantity on the left. In this way, causal equations differ from definitional equations in that they read more naturally right--to--left.  For example, from the electric field theory point of view, we say that the electric field at the location of a test charge is defined as the force $F$ divided by the charge $q$, $E = F/q$. 


Crucially, the context of each equation within a larger problem or argument is important for determining whether to interpret the equal sign as definitional or causal.

\subsection{Assignment (A)}
Although definitional and causal equations represent foundational physical relationships, it is sometimes necessary to temporarily associate concepts or variables to each other. We label these temporary relations as assignments with an operational articulation of ``let this equal that''. In the simplest cases, this form assigns numerical values to quantities (e.g. $t=4$) for use in solving problems or other manipulations. A more complex form is seen with symbolic assignments
\begin{equation} 
\label{eq4}
F{_{net}}=–kx-mg
\end{equation}
Discussed earlier, this equation encapsulates the idea that the net force on a mass hung from a spring is the sum of the gravitational and elastic forces. The net force is not always represented by this sum. Hence, this equation is not definitional, nor does the term $ –kx- mg$ \textit{cause} a net force. Rather $F_{net}$ and the sum $ –kx - mg$ may be used interchangeably for immediately subsequent calculations.

\subsection{Balancing (B)}
Dynamic equilibrium is a physical concept in which two (or more) quantities are in balance, numerically equivalent, and often directionally oppositional~\cite{Sherin2001}. When a mass hung from a spring reaches equilibrium and the net force is zero we write $kx=–mg$, indicating that the force from the spring $kx$ is equal and opposite to the gravitational force $mg$ with the symbol template $\msquare=\msquare$. The symbol template represents the structure of a mathematical expression without state the values or variables. Boxes demonstrate group of symbols (quantities or variables).

Balancing can be independent of direction, however, as in the conservation equation~(\ref{eq5}) which represents the balance between the flux of a vector field and the time rate of change of an associated density field.

\begin{equation} 
\label{eq5}
J=-\frac{\partial\rho}{\partial t}
\end{equation}

Unlike the previous categories, balancing equations may be read in either direction, as the equation does not emphasize or elevate one quantity over another. 

\subsection{Calculate (M)}
The final category identified is purely manipulative, indicating the result of a calculation. It can be thought of as equivalent to the use of a calculator button; a canonical example is $4+5=9$.

\section{Results}

\begin{figure*}[thb]
\includegraphics[width=170mm]{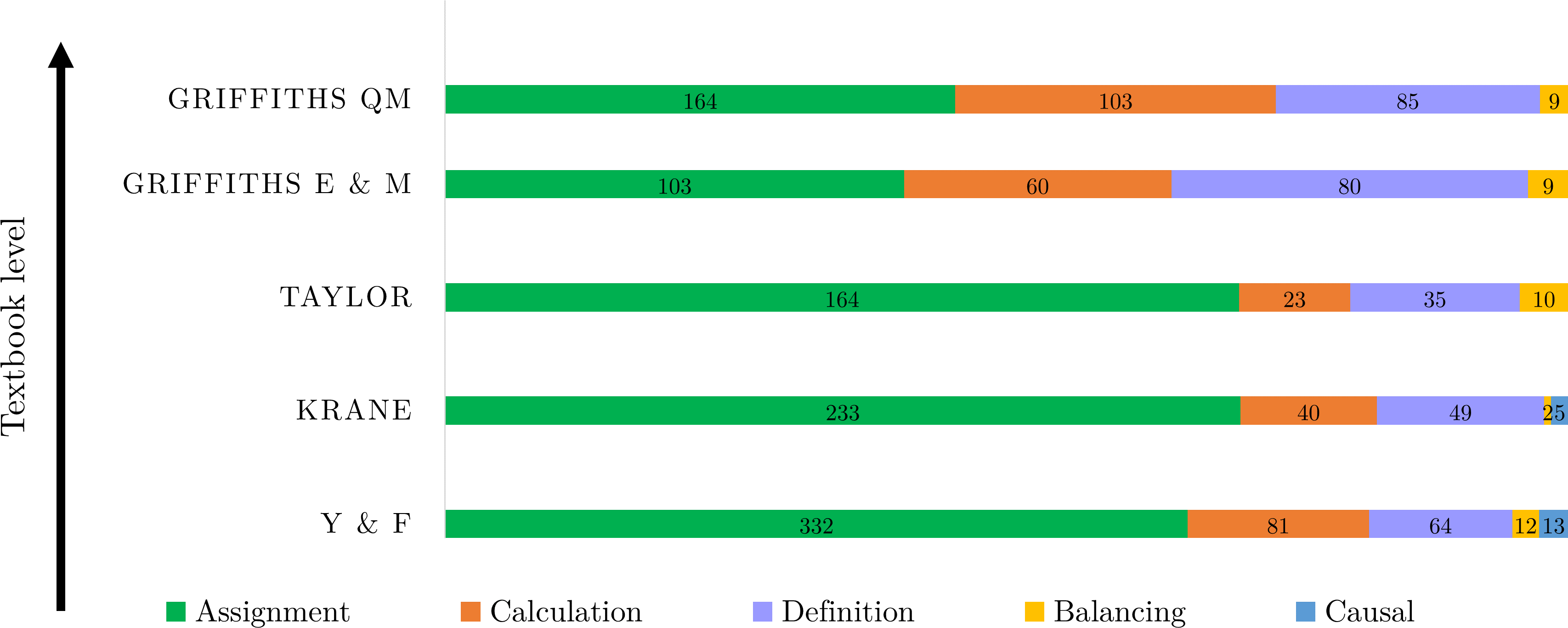}
\caption{Frequency of each equal sign category across 5 physics textbooks}  \label{Fig2-=sign}
\end{figure*}
1,676 separate equal signs were identified and coded in the 5 textbook chapters studied, an average of 335 per chapter. The distribution of usage by category for each chapter is shown in Figure \ref{Fig2-=sign}. In Figure \ref{Fig2-=sign}, textbooks are listed in order of increasing content level, from beginner (bottom) to most advanced (top). All bars are normalized to \text{100 \%}, with numbers overlaid to indicate the real numbers of codes in each category.

\begin{table}[b]
\centering
\caption{Summary of categories identified in textbooks, including operational articulation used to identify type, example and direction in which equations containing this type of sign are most easily read.}
\scalebox{0.93}{
\renewcommand{\arraystretch}{1.4}
\hskip-0.2cm\begin{tabular}{l|l|l|l}
\hline\hline
\multicolumn{1}{c|}{\textbf{Category}} & \multicolumn{1}{c|}{\textbf{Articulation}} & \multicolumn{1}{c|}{\textbf{Example}} & \multicolumn{1}{c}{\textbf{Direction}}\\ \hline
Definitional                         & ``Is defined as...''                          & $m=F/a$                                & Left--to--right                          \\ \hline
Causality                               & ``Leads to''                                & $a = F/m$                            & Right--to--left                          \\ \hline
Assignment                            & ``Let this = that''                         & $Y = c/2m$                           & Left-to-right                          \\ \hline
Balancing                             & ``This is balanced by...''                    & $kx = -mg$                            & Bidirectional                          \\ \hline
Calculate                           & ``The rest is just math...''                  & $4 + 5 = 9$                                      & Left--to--right \\ \hline

\end{tabular}}
\label{tab:2}
\end{table}

Introductory and intermediate textbooks (bottom three rows) show a higher proportion of assignment--type equals signs, with (on average) \text{69 \%} of all equal signs found to be of this type. These texts also have more example problems than advanced texts, and the quantitative nature of such problems as well as formulaic, step--by--step explanations contain significant portion of both the purely numerical (e.g. $t=5$) and symbolic (e.g. $F=mg$) assignments observed. Upper--level textbooks have a significantly smaller (average \text{43 \%}) portion of assignment equal signs. 

Surprisingly, advanced textbooks have twice the fraction (\text{26 \%} vs. \text{13 \%}) of signs classified as calculation. The complicated derivations found in upper-level textbooks involve a high amount of symbolic manipulation, and hence include a large number of equals signs of this type. The derivations also rely upon more carefully defined quantities, and so have a larger fraction of definitional equals signs. The upper-level textbooks also have a surprising dearth of causal equals signs, even controlling across content.

In addition to a shift in the frequency of the equal sign categories, there are also changes in the sub-type of assignment equal sign as the material becomes more advanced. Introductory textbooks have a roughly even distribution of usage between symbolic and numeric assignments, a consequence of the many worked problems with numbers given. Intermediate and advanced textbooks, however, use far greater proportions of symbolic assignments. The intermediate \textit{Mechanics} textbook~\cite{Taylor2005} had a 10:1 ratio of symbolic--to--numeric assignment equal signs, where the advanced \textit{Electricity \& Magnetism} book~\cite{Griffiths1999EM} had a 22:1 ratio. Even when sample problems are present in these texts, the use of numbers is discounted in favor of more abstract, symbolic representations.

\section{Conclusions}

A categorization scheme has been developed and validated internally for consistency among researchers as well as externally for resonance within the discipline. Five categories are identified, with symbolic and numeric sub-categories also appearing. Our categorization scheme supplements Sherin's symbolic forms~\cite{Sherin2001,Sherin2006}. Whereas Sherin ascribed meaning to entire equations, we argue that, at least in some equations, the meaning is mediated by the type of equal sign used. More broadly, we posit that the embedded conceptual meaning is contained specifically in the mathematical operators (symbols for addition, subtraction, multiplication, division, integration, differentiation, etc.) as these define relations between physics concepts. This meaning depends on the quantities being related (e.g. $F=ma$ has a different conceptual meaning than $F=mg$) and the difference is expressed in the relation, i.e. the operational symbols.

Understanding the equal sign as a relational symbol is more important in upper-level courses, where advanced problems are more symbolic than numeric. This requires an accurate perception about the relational meaning of the equals sign. This study is the first look at how undergraduate level physics textbooks communicate equal signs. Evidence indicates that introductory textbooks use more simple and operational types of equal signs, while advanced textbook incorporate a greater proportion of symbolic  assignments. More bridging between operational and relational forms is needed to develop in students a better understanding of these nuanced differences.

Drawing direct applications to instruction from this study would be premature, however; a valuable next step in this research would be a study connecting fundamental and applied research (e.g. curriculum development) with an eye on developing instructional goals. However, we speculate that helping instructors obtain a view of the equals sign as a relational symbol that has a different conceptual meaning might aid students in making connections between mathematics and physics.  As this distinction becomes more important the further students progress in their physics classes, attending to cultural meanings of the equal sign might help more advanced students take up ways of thinking like physicists\cite{Sayre2015Brief}.


We encourage instructors to provide opportunities such as collaborative problem solving for students to foster their reasoning in the classroom and engage them in conversations about the equal sign. As students work together and talk about  physics and mathematics formalism, the collaborative nature of these problem solving environments may help them pick up on and productively use physics cultural meanings in mathematical formalism.


Future work in this area could proceed along multiple lines. First, instructor discourse surrounding use of symbols during classroom practice could be investigated. Such work would identify how instructors attend to the conceptual meanings of symbols, with practical implications for instruction. Alternately, student articulations of meaning while solving problems could yield insight into their conceptual understanding and models of physics principles, analogous with recent work on symbolic forms, (e.g. Kuo \textit{et al.}),~\cite{kh2013}. Finally, physics \textit{practitioner} use (in, for example, research presentations) could be analyzed, similar to the work of Burton~\cite{Burton2000} to understand how articulations of concepts and meaning are used in communications between experts. Such articulations could then be compared to the articulations presented in instructional materials.

\section{Acknowledgments}
We would like to thank the Kansas State University Physics Education Research group (KSUPER) for their support and for helpful feedback. Three anonymous reviewers at ICLS gave us helpful feedback about an earlier version of this paper. We also thank Bahar Modir for their assistance with the IRR. This study is supported by the KSU Department of Physics, NSF grants DUE-1430967 and DUE-1317450 and the RIT Center for Advancing STEM Teaching, Learning\& Evaluation.


\begin{thebibliography}{66}%
\makeatletter
\providecommand \@ifxundefined [1]{%
 \@ifx{#1\undefined}
}%
\providecommand \@ifnum [1]{%
 \ifnum #1\expandafter \@firstoftwo
 \else \expandafter \@secondoftwo
 \fi
}%
\providecommand \@ifx [1]{%
 \ifx #1\expandafter \@firstoftwo
 \else \expandafter \@secondoftwo
 \fi
}%
\providecommand \natexlab [1]{#1}%
\providecommand \enquote  [1]{``#1''}%
\providecommand \bibnamefont  [1]{#1}%
\providecommand \bibfnamefont [1]{#1}%
\providecommand \citenamefont [1]{#1}%
\providecommand \href@noop [0]{\@secondoftwo}%
\providecommand \href [0]{\begingroup \@sanitize@url \@href}%
\providecommand \@href[1]{\@@startlink{#1}\@@href}%
\providecommand \@@href[1]{\endgroup#1\@@endlink}%
\providecommand \@sanitize@url [0]{\catcode `\\12\catcode `\$12\catcode
  `\&12\catcode `\#12\catcode `\^12\catcode `\_12\catcode `\%12\relax}%
\providecommand \@@startlink[1]{}%
\providecommand \@@endlink[0]{}%
\providecommand \url  [0]{\begingroup\@sanitize@url \@url }%
\providecommand \@url [1]{\endgroup\@href {#1}{\urlprefix }}%
\providecommand \urlprefix  [0]{URL }%
\providecommand \Eprint [0]{\href }%
\providecommand \doibase [0]{http://dx.doi.org/}%
\providecommand \selectlanguage [0]{\@gobble}%
\providecommand \bibinfo  [0]{\@secondoftwo}%
\providecommand \bibfield  [0]{\@secondoftwo}%
\providecommand \translation [1]{[#1]}%
\providecommand \BibitemOpen [0]{}%
\providecommand \bibitemStop [0]{}%
\providecommand \bibitemNoStop [0]{.\EOS\space}%
\providecommand \EOS [0]{\spacefactor3000\relax}%
\providecommand \BibitemShut  [1]{\csname bibitem#1\endcsname}%
\let\auto@bib@innerbib\@empty
\bibitem [{\citenamefont {J.~Knuth}\ \emph {et~al.}(2006)\citenamefont
  {J.~Knuth}, \citenamefont {Stephens}, \citenamefont {McNeil},\ and\
  \citenamefont {Alibali}}]{Knuth2006}%
  \BibitemOpen
  \bibfield  {author} {\bibinfo {author} {\bibfnamefont {E.}~\bibnamefont
  {J.~Knuth}}, \bibinfo {author} {\bibfnamefont {A.}~\bibnamefont {Stephens}},
  \bibinfo {author} {\bibfnamefont {N.}~\bibnamefont {McNeil}}, \ and\ \bibinfo
  {author} {\bibfnamefont {M.}~\bibnamefont {Alibali}},\ }\href {\doibase
  10.2307/30034852} {\bibfield  {journal} {\bibinfo  {journal} {J. Res. Math.
  Educ.}\ }\textbf {\bibinfo {volume} {37}},\ \bibinfo {pages} {297} (\bibinfo
  {year} {2006})}\BibitemShut {NoStop}%
\bibitem [{\citenamefont {Molina}\ and\ \citenamefont
  {Ambrose}(2006)}]{Molina2007}%
  \BibitemOpen
  \bibfield  {author} {\bibinfo {author} {\bibfnamefont {M.}~\bibnamefont
  {Molina}}\ and\ \bibinfo {author} {\bibfnamefont {R.}~\bibnamefont
  {Ambrose}},\ }\href@noop {} {\bibfield  {journal} {\bibinfo  {journal}
  {Teaching Children Mathematics}\ }\textbf {\bibinfo {volume} {13}},\ \bibinfo
  {pages} {111} (\bibinfo {year} {2006})}\BibitemShut {NoStop}%
\bibitem [{\citenamefont {Knuth}\ \emph {et~al.}(2008)\citenamefont {Knuth},
  \citenamefont {Alibali}, \citenamefont {Hattikudur}, \citenamefont {McNeil},\
  and\ \citenamefont {Stephens}}]{Knuth2008}%
  \BibitemOpen
  \bibfield  {author} {\bibinfo {author} {\bibfnamefont {E.}~\bibnamefont
  {Knuth}}, \bibinfo {author} {\bibfnamefont {M.}~\bibnamefont {Alibali}},
  \bibinfo {author} {\bibfnamefont {S.}~\bibnamefont {Hattikudur}}, \bibinfo
  {author} {\bibfnamefont {N.}~\bibnamefont {McNeil}}, \ and\ \bibinfo {author}
  {\bibfnamefont {A.}~\bibnamefont {Stephens}},\ }\href@noop {} {\bibfield
  {journal} {\bibinfo  {journal} {Mathematics Teaching in the Middle School}\
  }\textbf {\bibinfo {volume} {13}} (\bibinfo {year} {2008})}\BibitemShut
  {NoStop}%
\bibitem [{\citenamefont {Rittle-Johnson}\ \emph {et~al.}(2011)\citenamefont
  {Rittle-Johnson}, \citenamefont {Matthews}, \citenamefont {Taylor},\ and\
  \citenamefont {McEldoon}}]{RittleJohnson2011}%
  \BibitemOpen
  \bibfield  {author} {\bibinfo {author} {\bibfnamefont {B.}~\bibnamefont
  {Rittle-Johnson}}, \bibinfo {author} {\bibfnamefont {P.~G.}\ \bibnamefont
  {Matthews}}, \bibinfo {author} {\bibfnamefont {R.~S.}\ \bibnamefont
  {Taylor}}, \ and\ \bibinfo {author} {\bibfnamefont {K.~L.}\ \bibnamefont
  {McEldoon}},\ }\href@noop {} {\bibfield  {journal} {\bibinfo  {journal} {J.
  Educ. Psychol.}\ }\textbf {\bibinfo {volume} {103}},\ \bibinfo {pages} {85}
  (\bibinfo {year} {2011})}\BibitemShut {NoStop}%
\bibitem [{\citenamefont {Sherman}\ and\ \citenamefont
  {Bisanz}(2009)}]{Sherman2009}%
  \BibitemOpen
  \bibfield  {author} {\bibinfo {author} {\bibfnamefont {J.}~\bibnamefont
  {Sherman}}\ and\ \bibinfo {author} {\bibfnamefont {J.}~\bibnamefont
  {Bisanz}},\ }\href@noop {} {\bibfield  {journal} {\bibinfo  {journal}
  {Journal of Educational Psychology}\ }\textbf {\bibinfo {volume} {101}},\
  \bibinfo {pages} {88} (\bibinfo {year} {2009})}\BibitemShut {NoStop}%
\bibitem [{\citenamefont {Stephens}\ \emph {et~al.}(2013)\citenamefont
  {Stephens}, \citenamefont {Knuth}, \citenamefont {Blanton}, \citenamefont
  {Isler-Baykal}, \citenamefont {Gardiner},\ and\ \citenamefont
  {Marum}}]{Stephens2013}%
  \BibitemOpen
  \bibfield  {author} {\bibinfo {author} {\bibfnamefont {A.}~\bibnamefont
  {Stephens}}, \bibinfo {author} {\bibfnamefont {E.}~\bibnamefont {Knuth}},
  \bibinfo {author} {\bibfnamefont {M.}~\bibnamefont {Blanton}}, \bibinfo
  {author} {\bibfnamefont {I.}~\bibnamefont {Isler-Baykal}}, \bibinfo {author}
  {\bibfnamefont {A.}~\bibnamefont {Gardiner}}, \ and\ \bibinfo {author}
  {\bibfnamefont {T.}~\bibnamefont {Marum}},\ }\href {\doibase
  10.1016/j.jmathb.2013.02.001} {\bibfield  {journal} {\bibinfo  {journal} {The
  Journal of Mathematical Behavior}\ }\textbf {\bibinfo {volume} {32}},\
  \bibinfo {pages} {173–} (\bibinfo {year} {2013})}\BibitemShut {NoStop}%
\bibitem [{\citenamefont {Renwick}(1932)}]{Renwick1932}%
  \BibitemOpen
  \bibfield  {author} {\bibinfo {author} {\bibfnamefont {E.~M.}\ \bibnamefont
  {Renwick}},\ }\href@noop {} {\bibfield  {journal} {\bibinfo  {journal} {Br.
  J. Educ. Psychol.}\ }\textbf {\bibinfo {volume} {2}},\ \bibinfo {pages} {173}
  (\bibinfo {year} {1932})}\BibitemShut {NoStop}%
\bibitem [{\citenamefont {Denmark}\ and\ \citenamefont
  {Barco}(1976)}]{Denmak1976}%
  \BibitemOpen
  \bibfield  {author} {\bibinfo {author} {\bibfnamefont {T.}~\bibnamefont
  {Denmark}}\ and\ \bibinfo {author} {\bibfnamefont {J.}~\bibnamefont {Barco},
  \bibfnamefont {E.and~Voran}},\ }\href@noop {} {\bibfield  {journal} {\bibinfo
   {journal} {PMDC Technical Report}\ } (\bibinfo {year} {1976})}\BibitemShut
  {NoStop}%
\bibitem [{\citenamefont {Behr}\ \emph {et~al.}(1976)\citenamefont {Behr},
  \citenamefont {Erlwanger},\ and\ \citenamefont {Nichols}}]{Behr1976}%
  \BibitemOpen
  \bibfield  {author} {\bibinfo {author} {\bibfnamefont {M.}~\bibnamefont
  {Behr}}, \bibinfo {author} {\bibfnamefont {S.}~\bibnamefont {Erlwanger}}, \
  and\ \bibinfo {author} {\bibfnamefont {E.}~\bibnamefont {Nichols}},\
  }\href@noop {} {\bibfield  {journal} {\bibinfo  {journal} {PMDC Technical
  Report}\ } (\bibinfo {year} {1976})}\BibitemShut {NoStop}%
\bibitem [{\citenamefont {Ginsburg}(1977)}]{ginsburg1977}%
  \BibitemOpen
  \bibfield  {author} {\bibinfo {author} {\bibfnamefont {H.}~\bibnamefont
  {Ginsburg}},\ }\href@noop {} {\emph {\bibinfo {title} {Children's arithmetic:
  The learning process.}}}\ (\bibinfo  {publisher} {D. van Nostrand},\ \bibinfo
  {year} {1977})\BibitemShut {NoStop}%
\bibitem [{\citenamefont {Behr}\ \emph {et~al.}(1980)\citenamefont {Behr},
  \citenamefont {Erlwanger},\ and\ \citenamefont {Nichols}}]{Behr1980}%
  \BibitemOpen
  \bibfield  {author} {\bibinfo {author} {\bibfnamefont {M.}~\bibnamefont
  {Behr}}, \bibinfo {author} {\bibfnamefont {S.}~\bibnamefont {Erlwanger}}, \
  and\ \bibinfo {author} {\bibfnamefont {E.}~\bibnamefont {Nichols}},\
  }\href@noop {} {\bibfield  {journal} {\bibinfo  {journal} {Mathematics
  Teaching}\ }\textbf {\bibinfo {volume} {92}},\ \bibinfo {pages} {13}
  (\bibinfo {year} {1980})}\BibitemShut {NoStop}%
\bibitem [{\citenamefont {Kieran}(1981)}]{Kieran1981}%
  \BibitemOpen
  \bibfield  {author} {\bibinfo {author} {\bibfnamefont {C.}~\bibnamefont
  {Kieran}},\ }\href@noop {} {\bibfield  {journal} {\bibinfo  {journal} {Educ.
  Stud. Math.}\ }\textbf {\bibinfo {volume} {12}},\ \bibinfo {pages} {317}
  (\bibinfo {year} {1981})}\BibitemShut {NoStop}%
\bibitem [{\citenamefont {Baroody}\ and\ \citenamefont
  {Ginsburg}(1982)}]{BaroodyGinsburg1982}%
  \BibitemOpen
  \bibfield  {author} {\bibinfo {author} {\bibfnamefont {A.~J.}\ \bibnamefont
  {Baroody}}\ and\ \bibinfo {author} {\bibfnamefont {H.}~\bibnamefont
  {Ginsburg}},\ }\href@noop {} {\bibfield  {journal} {\bibinfo  {journal}
  {Cogn. and Instr.}\ }\textbf {\bibinfo {volume} {24}},\ \bibinfo {pages}
  {367} (\bibinfo {year} {1982})}\BibitemShut {NoStop}%
\bibitem [{\citenamefont {S{\'a}enz-Ludlow}\ and\ \citenamefont
  {Walgamuth}(1998)}]{LudlowWalgamuth1998}%
  \BibitemOpen
  \bibfield  {author} {\bibinfo {author} {\bibfnamefont {A.}~\bibnamefont
  {S{\'a}enz-Ludlow}}\ and\ \bibinfo {author} {\bibfnamefont {C.}~\bibnamefont
  {Walgamuth}},\ }\href {\doibase 10.1023/A:1003086304201} {\bibfield
  {journal} {\bibinfo  {journal} {Educ. Stud. Math.}\ }\textbf {\bibinfo
  {volume} {35}},\ \bibinfo {pages} {153} (\bibinfo {year} {1998})}\BibitemShut
  {NoStop}%
\bibitem [{\citenamefont {Oksuz}(2007)}]{Oksuz2007}%
  \BibitemOpen
  \bibfield  {author} {\bibinfo {author} {\bibfnamefont {C.}~\bibnamefont
  {Oksuz}},\ }\href@noop {} {\bibfield  {journal} {\bibinfo  {journal}
  {International Journal for Mathematics Teaching and Learning [electronic
  only]}\ }\textbf {\bibinfo {volume} {2007}},\ \bibinfo {pages} {20} (\bibinfo
  {year} {2007})}\BibitemShut {NoStop}%
\bibitem [{\citenamefont {Noonan}\ and\ \citenamefont
  {Curtis}(2018)}]{Noonan2014}%
  \BibitemOpen
  \bibfield  {author} {\bibinfo {author} {\bibfnamefont {H.}~\bibnamefont
  {Noonan}}\ and\ \bibinfo {author} {\bibfnamefont {B.}~\bibnamefont
  {Curtis}},\ }in\ \href@noop {} {\emph {\bibinfo {booktitle} {The Stanford
  encyclopedia of philosophy}}},\ \bibinfo {editor} {edited by\ \bibinfo
  {editor} {\bibfnamefont {E.~N.}\ \bibnamefont {Zalta}}}\ (\bibinfo
  {publisher} {Metaphysics Research Lab, Stanford University},\ \bibinfo {year}
  {2018})\ \bibinfo {edition} {summer 2018}\ ed.\BibitemShut {Stop}%
\bibitem [{\citenamefont {Byrd}\ \emph {et~al.}(2015)\citenamefont {Byrd},
  \citenamefont {McNeil}, \citenamefont {Chesney},\ and\ \citenamefont
  {Matthews}}]{ByrdMcNeilChesneyMatthews2015}%
  \BibitemOpen
  \bibfield  {author} {\bibinfo {author} {\bibfnamefont {C.~E.}\ \bibnamefont
  {Byrd}}, \bibinfo {author} {\bibfnamefont {N.~M.}\ \bibnamefont {McNeil}},
  \bibinfo {author} {\bibfnamefont {D.~L.}\ \bibnamefont {Chesney}}, \ and\
  \bibinfo {author} {\bibfnamefont {P.~G.}\ \bibnamefont {Matthews}},\ }\href
  {\doibase https://doi.org/10.1016/j.lindif.2015.01.001} {\bibfield  {journal}
  {\bibinfo  {journal} {Learn Individ. Differ.}\ }\textbf {\bibinfo {volume}
  {38}},\ \bibinfo {pages} {61} (\bibinfo {year} {2015})}\BibitemShut {NoStop}%
\bibitem [{\citenamefont {P.~Falkner}\ \emph {et~al.}(1999)\citenamefont
  {P.~Falkner}, \citenamefont {Levi},\ and\ \citenamefont
  {P.~Carpenter}}]{Falkner1999}%
  \BibitemOpen
  \bibfield  {author} {\bibinfo {author} {\bibfnamefont {K.}~\bibnamefont
  {P.~Falkner}}, \bibinfo {author} {\bibfnamefont {L.}~\bibnamefont {Levi}}, \
  and\ \bibinfo {author} {\bibfnamefont {T.}~\bibnamefont {P.~Carpenter}},\
  }\href@noop {} {\bibfield  {journal} {\bibinfo  {journal} {Teach. Child.
  Math.}\ }\textbf {\bibinfo {volume} {6}},\ \bibinfo {pages} {232} (\bibinfo
  {year} {1999})}\BibitemShut {NoStop}%
\bibitem [{\citenamefont {Heller}\ \emph
  {et~al.}(1992{\natexlab{a}})\citenamefont {Heller}, \citenamefont {Keith},\
  and\ \citenamefont {Anderson}}]{Hellerp11991}%
  \BibitemOpen
  \bibfield  {author} {\bibinfo {author} {\bibfnamefont {P.}~\bibnamefont
  {Heller}}, \bibinfo {author} {\bibfnamefont {R.}~\bibnamefont {Keith}}, \
  and\ \bibinfo {author} {\bibfnamefont {S.}~\bibnamefont {Anderson}},\ }\href
  {\doibase 10.1119/1.17117} {\bibfield  {journal} {\bibinfo  {journal} {Am. J.
  Phys.}\ }\textbf {\bibinfo {volume} {60}},\ \bibinfo {pages} {627} (\bibinfo
  {year} {1992}{\natexlab{a}})}\BibitemShut {NoStop}%
\bibitem [{\citenamefont {Heller}\ and\ \citenamefont
  {Hollabaugh}(1992)}]{Hellerp21991}%
  \BibitemOpen
  \bibfield  {author} {\bibinfo {author} {\bibfnamefont {P.}~\bibnamefont
  {Heller}}\ and\ \bibinfo {author} {\bibfnamefont {M.}~\bibnamefont
  {Hollabaugh}},\ }\href {\doibase 10.1119/1.17118} {\bibfield  {journal}
  {\bibinfo  {journal} {Am. J. Phys.}\ }\textbf {\bibinfo {volume} {60}},\
  \bibinfo {pages} {637} (\bibinfo {year} {1992})}\BibitemShut {NoStop}%
\bibitem [{\citenamefont {Gabel}(1994)}]{Gabel1994}%
  \BibitemOpen
  \bibfield  {author} {\bibinfo {author} {\bibfnamefont {D.}~\bibnamefont
  {Gabel}},\ }\href@noop {} {\emph {\bibinfo {title} {Handbook of research on
  science teaching and learning}}}\ (\bibinfo  {publisher} {New York:
  Macmillan},\ \bibinfo {year} {1994})\BibitemShut {NoStop}%
\bibitem [{\citenamefont {Thacker}\ \emph {et~al.}(1994)\citenamefont
  {Thacker}, \citenamefont {Kim}, \citenamefont {Trefz},\ and\ \citenamefont
  {Lea}}]{Thacker1994}%
  \BibitemOpen
  \bibfield  {author} {\bibinfo {author} {\bibfnamefont {B.}~\bibnamefont
  {Thacker}}, \bibinfo {author} {\bibfnamefont {E.}~\bibnamefont {Kim}},
  \bibinfo {author} {\bibfnamefont {K.}~\bibnamefont {Trefz}}, \ and\ \bibinfo
  {author} {\bibfnamefont {S.~M.}\ \bibnamefont {Lea}},\ }\href {\doibase
  10.1119/1.17480} {\bibfield  {journal} {\bibinfo  {journal} {Am. J. Phys.}\
  }\textbf {\bibinfo {volume} {62}},\ \bibinfo {pages} {627} (\bibinfo {year}
  {1994})}\BibitemShut {NoStop}%
\bibitem [{\citenamefont {Hsu}\ \emph {et~al.}(2004)\citenamefont {Hsu},
  \citenamefont {Brewe}, \citenamefont {Foster},\ and\ \citenamefont
  {Harper}}]{Hsu2004}%
  \BibitemOpen
  \bibfield  {author} {\bibinfo {author} {\bibfnamefont {L.}~\bibnamefont
  {Hsu}}, \bibinfo {author} {\bibfnamefont {E.}~\bibnamefont {Brewe}}, \bibinfo
  {author} {\bibfnamefont {T.~M.}\ \bibnamefont {Foster}}, \ and\ \bibinfo
  {author} {\bibfnamefont {K.~A.}\ \bibnamefont {Harper}},\ }\href {\doibase
  10.1119/1.1763175} {\bibfield  {journal} {\bibinfo  {journal} {Am. J. Phys.}\
  }\textbf {\bibinfo {volume} {72}},\ \bibinfo {pages} {1147} (\bibinfo {year}
  {2004})}\BibitemShut {NoStop}%
\bibitem [{\citenamefont {Meltzer}(2005)}]{Meltzer2005}%
  \BibitemOpen
  \bibfield  {author} {\bibinfo {author} {\bibfnamefont {D.~E.}\ \bibnamefont
  {Meltzer}},\ }\href {\doibase 10.1119/1.1862636} {\bibfield  {journal}
  {\bibinfo  {journal} {Am. J. Phys.}\ }\textbf {\bibinfo {volume} {73}},\
  \bibinfo {pages} {463} (\bibinfo {year} {2005})}\BibitemShut {NoStop}%
\bibitem [{\citenamefont {Fredlund}\ \emph {et~al.}(2014)\citenamefont
  {Fredlund}, \citenamefont {Linder}, \citenamefont {Airey},\ and\
  \citenamefont {Linder}}]{Fredlund2014}%
  \BibitemOpen
  \bibfield  {author} {\bibinfo {author} {\bibfnamefont {T.}~\bibnamefont
  {Fredlund}}, \bibinfo {author} {\bibfnamefont {C.}~\bibnamefont {Linder}},
  \bibinfo {author} {\bibfnamefont {J.}~\bibnamefont {Airey}}, \ and\ \bibinfo
  {author} {\bibfnamefont {A.}~\bibnamefont {Linder}},\ }\href@noop {}
  {\bibfield  {journal} {\bibinfo  {journal} {Phys. Rev. ST Phys. Educ. Res.}\
  }\textbf {\bibinfo {volume} {10}},\ \bibinfo {pages} {020129} (\bibinfo
  {year} {2014})}\BibitemShut {NoStop}%
\bibitem [{\citenamefont {Scherr}(2008)}]{Scherr2008}%
  \BibitemOpen
  \bibfield  {author} {\bibinfo {author} {\bibfnamefont {R.~E.}\ \bibnamefont
  {Scherr}},\ }\href {\doibase 10.1103/PhysRevSTPER.4.010101} {\bibfield
  {journal} {\bibinfo  {journal} {Phys. Rev. ST Phys. Educ. Res.}\ }\textbf
  {\bibinfo {volume} {4}},\ \bibinfo {pages} {010101} (\bibinfo {year}
  {2008})}\BibitemShut {NoStop}%
\bibitem [{\citenamefont {Aberg-Bengtsson}\ and\ \citenamefont
  {Ottosson}(2006)}]{Ottosson2006}%
  \BibitemOpen
  \bibfield  {author} {\bibinfo {author} {\bibfnamefont {L.}~\bibnamefont
  {Aberg-Bengtsson}}\ and\ \bibinfo {author} {\bibfnamefont {T.}~\bibnamefont
  {Ottosson}},\ }\href {\doibase 10.1002/tea.20087} {\bibfield  {journal}
  {\bibinfo  {journal} {J. of Research in Sci. Teaching}\ }\textbf {\bibinfo
  {volume} {43}},\ \bibinfo {pages} {43} (\bibinfo {year} {2006})}\BibitemShut
  {NoStop}%
\bibitem [{\citenamefont {Rosengrant}\ \emph {et~al.}(2009)\citenamefont
  {Rosengrant}, \citenamefont {Van~Heuvelen},\ and\ \citenamefont
  {Etkina}}]{Rosengrant2009}%
  \BibitemOpen
  \bibfield  {author} {\bibinfo {author} {\bibfnamefont {D.}~\bibnamefont
  {Rosengrant}}, \bibinfo {author} {\bibfnamefont {A.}~\bibnamefont
  {Van~Heuvelen}}, \ and\ \bibinfo {author} {\bibfnamefont {E.}~\bibnamefont
  {Etkina}},\ }\href {\doibase 10.1103/PhysRevSTPER.5.010108} {\bibfield
  {journal} {\bibinfo  {journal} {Phys. Rev. ST Phys. Educ. Res.}\ }\textbf
  {\bibinfo {volume} {5}},\ \bibinfo {pages} {010108} (\bibinfo {year}
  {2009})}\BibitemShut {NoStop}%
\bibitem [{\citenamefont {Fredlund}\ \emph {et~al.}(2012)\citenamefont
  {Fredlund}, \citenamefont {Airey},\ and\ \citenamefont
  {Linder}}]{Fredlund2012}%
  \BibitemOpen
  \bibfield  {author} {\bibinfo {author} {\bibfnamefont {T.}~\bibnamefont
  {Fredlund}}, \bibinfo {author} {\bibfnamefont {J.}~\bibnamefont {Airey}}, \
  and\ \bibinfo {author} {\bibfnamefont {C.}~\bibnamefont {Linder}},\ }\href
  {\doibase 10.1088/0143-0807/33/3/657} {\bibfield  {journal} {\bibinfo
  {journal} {Eur. J. Phys.}\ }\textbf {\bibinfo {volume} {33}},\ \bibinfo
  {pages} {657} (\bibinfo {year} {2012})}\BibitemShut {NoStop}%
\bibitem [{\citenamefont {Christensen}\ and\ \citenamefont
  {Thompson}(2012)}]{christensen2012}%
  \BibitemOpen
  \bibfield  {author} {\bibinfo {author} {\bibfnamefont {W.~M.}\ \bibnamefont
  {Christensen}}\ and\ \bibinfo {author} {\bibfnamefont {J.~R.}\ \bibnamefont
  {Thompson}},\ }\href@noop {} {\bibfield  {journal} {\bibinfo  {journal}
  {Phys. Rev. ST Phys. Educ. Res.}\ }\textbf {\bibinfo {volume} {8}},\ \bibinfo
  {pages} {023101} (\bibinfo {year} {2012})}\BibitemShut {NoStop}%
\bibitem [{\citenamefont {Sherin}(2001)}]{Sherin2001}%
  \BibitemOpen
  \bibfield  {author} {\bibinfo {author} {\bibfnamefont {B.~L.}\ \bibnamefont
  {Sherin}},\ }\href@noop {} {\bibfield  {journal} {\bibinfo  {journal} {Cogn.
  Instr.}\ }\textbf {\bibinfo {volume} {19}},\ \bibinfo {pages} {479} (\bibinfo
  {year} {2001})}\BibitemShut {NoStop}%
\bibitem [{\citenamefont {Sherin}(2006)}]{Sherin2006}%
  \BibitemOpen
  \bibfield  {author} {\bibinfo {author} {\bibfnamefont {B.~L.}\ \bibnamefont
  {Sherin}},\ }\href@noop {} {\bibfield  {journal} {\bibinfo  {journal} {J.
  Res. Sci. Teach.}\ }\textbf {\bibinfo {volume} {43}},\ \bibinfo {pages} {535}
  (\bibinfo {year} {2006})}\BibitemShut {NoStop}%
\bibitem [{\citenamefont {Ragout De~Lozano}\ and\ \citenamefont
  {Cardenas}(2002)}]{Ragout2002}%
  \BibitemOpen
  \bibfield  {author} {\bibinfo {author} {\bibfnamefont {S.}~\bibnamefont
  {Ragout De~Lozano}}\ and\ \bibinfo {author} {\bibfnamefont {M.}~\bibnamefont
  {Cardenas}},\ }\href {\doibase 10.1023/A:1019643420896} {\bibfield  {journal}
  {\bibinfo  {journal} {Sci. Educ.}\ }\textbf {\bibinfo {volume} {11}},\
  \bibinfo {pages} {589} (\bibinfo {year} {2002})}\BibitemShut {NoStop}%
\bibitem [{\citenamefont {Domert}\ \emph {et~al.}(2007)\citenamefont {Domert},
  \citenamefont {Airey}, \citenamefont {Linder},\ and\ \citenamefont
  {Lippmann~Kung}}]{Domert2007}%
  \BibitemOpen
  \bibfield  {author} {\bibinfo {author} {\bibfnamefont {D.}~\bibnamefont
  {Domert}}, \bibinfo {author} {\bibfnamefont {J.}~\bibnamefont {Airey}},
  \bibinfo {author} {\bibfnamefont {C.}~\bibnamefont {Linder}}, \ and\ \bibinfo
  {author} {\bibfnamefont {R.}~\bibnamefont {Lippmann~Kung}},\ }\href
  {https://www.journals.uio.no/index.php/nordina/article/view/389} {\bibfield
  {journal} {\bibinfo  {journal} {NorDiNa: Nordic Studies in Science
  Education}\ }\textbf {\bibinfo {volume} {3}},\ \bibinfo {pages} {15}
  (\bibinfo {year} {2007})}\BibitemShut {NoStop}%
\bibitem [{\citenamefont {Bing}\ and\ \citenamefont {Redish}(2009)}]{bing2009}%
  \BibitemOpen
  \bibfield  {author} {\bibinfo {author} {\bibfnamefont {T.~J.}\ \bibnamefont
  {Bing}}\ and\ \bibinfo {author} {\bibfnamefont {E.~F.}\ \bibnamefont
  {Redish}},\ }\href@noop {} {\bibfield  {journal} {\bibinfo  {journal} {Phys.
  Rev. ST Phys. Educ. Res.}\ }\textbf {\bibinfo {volume} {5}},\ \bibinfo
  {pages} {020108} (\bibinfo {year} {2009})}\BibitemShut {NoStop}%
\bibitem [{\citenamefont {Brookes}(2006)}]{Brookes2006}%
  \BibitemOpen
  \bibfield  {author} {\bibinfo {author} {\bibfnamefont {D.}~\bibnamefont
  {Brookes}},\ }\href@noop {} {\bibfield  {journal} {\bibinfo  {journal}
  {Unpublished PhD thesis, Rutgers, New Brunswick}\ } (\bibinfo {year}
  {2006})}\BibitemShut {NoStop}%
\bibitem [{\citenamefont {Airey}\ and\ \citenamefont
  {Linder}(2006)}]{Linder2006}%
  \BibitemOpen
  \bibfield  {author} {\bibinfo {author} {\bibfnamefont {J.}~\bibnamefont
  {Airey}}\ and\ \bibinfo {author} {\bibfnamefont {C.}~\bibnamefont {Linder}},\
  }\href {\doibase 10.1088/0143-0807/27/3/009} {\bibfield  {journal} {\bibinfo
  {journal} {Eur. J. Phys}\ }\textbf {\bibinfo {volume} {27}},\ \bibinfo
  {pages} {553 } (\bibinfo {year} {2006})}\BibitemShut {NoStop}%
\bibitem [{\citenamefont {Linder}\ \emph {et~al.}(2011)\citenamefont {Linder},
  \citenamefont {\"{O}stman}, \citenamefont {Roberts}, \citenamefont {Wickman},
  \citenamefont {Ericksen},\ and\ \citenamefont {MacKinnon}}]{Linder2011}%
  \BibitemOpen
  \bibfield  {author} {\bibinfo {author} {\bibfnamefont {C.}~\bibnamefont
  {Linder}}, \bibinfo {author} {\bibfnamefont {L.}~\bibnamefont {\"{O}stman}},
  \bibinfo {author} {\bibfnamefont {D.}~\bibnamefont {Roberts}}, \bibinfo
  {author} {\bibfnamefont {P.}~\bibnamefont {Wickman}}, \bibinfo {author}
  {\bibfnamefont {G.}~\bibnamefont {Ericksen}}, \ and\ \bibinfo {author}
  {\bibfnamefont {A.}~\bibnamefont {MacKinnon}},\ }\href {\doibase
  https://doi.org/10.4324/9780203843284} {\emph {\bibinfo {title} {Exploring
  the landscape of scientific literacy.}}}\ (\bibinfo  {publisher} {New York},\
  \bibinfo {year} {2011})\BibitemShut {NoStop}%
\bibitem [{\citenamefont {Huffman}(1998)}]{Huffman1997}%
  \BibitemOpen
  \bibfield  {author} {\bibinfo {author} {\bibfnamefont {D.}~\bibnamefont
  {Huffman}},\ }\href@noop {} {\bibfield  {journal} {\bibinfo  {journal} {J.
  Res. Sci. Teaching}\ }\textbf {\bibinfo {volume} {34}},\ \bibinfo {pages}
  {551} (\bibinfo {year} {1998})}\BibitemShut {NoStop}%
\bibitem [{\citenamefont {Van~Heuvelen}(1991)}]{VanHeuvelen1991b}%
  \BibitemOpen
  \bibfield  {author} {\bibinfo {author} {\bibfnamefont {A.}~\bibnamefont
  {Van~Heuvelen}},\ }\href {\doibase 10.1119/1.16668} {\bibfield  {journal}
  {\bibinfo  {journal} {Am. J. Phys.}\ }\textbf {\bibinfo {volume} {59}},\
  \bibinfo {pages} {898} (\bibinfo {year} {1991})}\BibitemShut {NoStop}%
\bibitem [{\citenamefont {Heller}\ \emph
  {et~al.}(1992{\natexlab{b}})\citenamefont {Heller}, \citenamefont {Keith},\
  and\ \citenamefont {Anderson}}]{Heller1992a}%
  \BibitemOpen
  \bibfield  {author} {\bibinfo {author} {\bibfnamefont {P.}~\bibnamefont
  {Heller}}, \bibinfo {author} {\bibfnamefont {R.}~\bibnamefont {Keith}}, \
  and\ \bibinfo {author} {\bibfnamefont {S.}~\bibnamefont {Anderson}},\
  }\href@noop {} {\bibfield  {journal} {\bibinfo  {journal} {Am. J. Phys.}\
  }\textbf {\bibinfo {volume} {60}},\ \bibinfo {pages} {627} (\bibinfo {year}
  {1992}{\natexlab{b}})}\BibitemShut {NoStop}%
\bibitem [{\citenamefont {Walsh}\ \emph {et~al.}(2007)\citenamefont {Walsh},
  \citenamefont {Howard},\ and\ \citenamefont {Bowe}}]{Walsh}%
  \BibitemOpen
  \bibfield  {author} {\bibinfo {author} {\bibfnamefont {L.~N.}\ \bibnamefont
  {Walsh}}, \bibinfo {author} {\bibfnamefont {R.~G.}\ \bibnamefont {Howard}}, \
  and\ \bibinfo {author} {\bibfnamefont {B.}~\bibnamefont {Bowe}},\ }\href
  {\doibase 10.1103/PhysRevSTPER.3.020108} {\bibfield  {journal} {\bibinfo
  {journal} {Phys. Rev. ST Phys. Educ. Res.}\ }\textbf {\bibinfo {volume}
  {3}},\ \bibinfo {pages} {020108} (\bibinfo {year} {2007})}\BibitemShut
  {NoStop}%
\bibitem [{\citenamefont {Redish}\ and\ \citenamefont
  {Smith}(2008)}]{Redish2008LookingEngineers}%
  \BibitemOpen
  \bibfield  {author} {\bibinfo {author} {\bibfnamefont {E.~F.}\ \bibnamefont
  {Redish}}\ and\ \bibinfo {author} {\bibfnamefont {K.~A.}\ \bibnamefont
  {Smith}},\ }\href@noop {} {\bibfield  {journal} {\bibinfo  {journal} {J. Eng.
  Educ.}\ }\textbf {\bibinfo {volume} {97}},\ \bibinfo {pages} {295} (\bibinfo
  {year} {2008})}\BibitemShut {NoStop}%
\bibitem [{\citenamefont {Reif}(2008)}]{Reif2008ApplyingDomains}%
  \BibitemOpen
  \bibfield  {author} {\bibinfo {author} {\bibfnamefont {F.}~\bibnamefont
  {Reif}},\ }\href@noop {} {\emph {\bibinfo {title} {Applying cognitive science
  to education: Thinking and learning in scientific and other complex
  domains}}}\ (\bibinfo  {publisher} {MIT press},\ \bibinfo {year}
  {2008})\BibitemShut {NoStop}%
\bibitem [{\citenamefont {Kuo}\ \emph {et~al.}(2013)\citenamefont {Kuo},
  \citenamefont {Hull}, \citenamefont {Gupta},\ and\ \citenamefont
  {Elby}}]{kh2013}%
  \BibitemOpen
  \bibfield  {author} {\bibinfo {author} {\bibfnamefont {E.}~\bibnamefont
  {Kuo}}, \bibinfo {author} {\bibfnamefont {M.}~\bibnamefont {Hull}}, \bibinfo
  {author} {\bibfnamefont {A.}~\bibnamefont {Gupta}}, \ and\ \bibinfo {author}
  {\bibfnamefont {A.}~\bibnamefont {Elby}},\ }\href {\doibase
  10.1002/sce.21043} {\bibfield  {journal} {\bibinfo  {journal} {Sci. Educ.}\
  }\textbf {\bibinfo {volume} {97}},\ \bibinfo {pages} {32} (\bibinfo {year}
  {2013})}\BibitemShut {NoStop}%
\bibitem [{\citenamefont {Leonard}\ \emph {et~al.}(1996)\citenamefont
  {Leonard}, \citenamefont {Dufresne},\ and\ \citenamefont
  {Mestre}}]{Leonard1996}%
  \BibitemOpen
  \bibfield  {author} {\bibinfo {author} {\bibfnamefont {W.~J.}\ \bibnamefont
  {Leonard}}, \bibinfo {author} {\bibfnamefont {R.~J.}\ \bibnamefont
  {Dufresne}}, \ and\ \bibinfo {author} {\bibfnamefont {J.}~\bibnamefont
  {Mestre}},\ }\href {\doibase 10.1119/1.18409} {\bibfield  {journal} {\bibinfo
   {journal} {Am. J. Phys.}\ }\textbf {\bibinfo {volume} {64}},\ \bibinfo
  {pages} {1495} (\bibinfo {year} {1996})}\BibitemShut {NoStop}%
\bibitem [{\citenamefont {{Mazur}}(1998)}]{Mazur1998}%
  \BibitemOpen
  \bibfield  {author} {\bibinfo {author} {\bibfnamefont {E.}~\bibnamefont
  {{Mazur}}},\ }in\ \href@noop {} {\emph {\bibinfo {booktitle} {American
  Astronomical Society Meeting Abstracts}}},\ \bibinfo {series} {Bulletin of
  the American Astronomical Society}, Vol.~\bibinfo {volume} {30}\ (\bibinfo
  {year} {1998})\ p.\ \bibinfo {pages} {1331}\BibitemShut {NoStop}%
\bibitem [{\citenamefont {Redish}(2005)}]{Redish2005}%
  \BibitemOpen
  \bibfield  {author} {\bibinfo {author} {\bibfnamefont {E.~F.}\ \bibnamefont
  {Redish}},\ }\href@noop {} {\bibfield  {journal} {\bibinfo  {journal} {to be
  published in Proceedings of the Conference, World View on Physics Education
  in 2005, Focusing on Change, Delhi}\ } (\bibinfo {year} {2005})}\BibitemShut
  {NoStop}%
\bibitem [{\citenamefont {Uhden}\ \emph {et~al.}(2012)\citenamefont {Uhden},
  \citenamefont {Karam}, \citenamefont {Pietrocola},\ and\ \citenamefont
  {Pospiech}}]{uhden2012}%
  \BibitemOpen
  \bibfield  {author} {\bibinfo {author} {\bibfnamefont {O.}~\bibnamefont
  {Uhden}}, \bibinfo {author} {\bibfnamefont {R.}~\bibnamefont {Karam}},
  \bibinfo {author} {\bibfnamefont {M.}~\bibnamefont {Pietrocola}}, \ and\
  \bibinfo {author} {\bibfnamefont {G.}~\bibnamefont {Pospiech}},\ }\href@noop
  {} {\bibfield  {journal} {\bibinfo  {journal} {Science {\&} Education}\
  }\textbf {\bibinfo {volume} {21}},\ \bibinfo {pages} {485} (\bibinfo {year}
  {2012})}\BibitemShut {NoStop}%
\bibitem [{\citenamefont {Freudenthal}(1973)}]{Freudenthal1973}%
  \BibitemOpen
  \bibfield  {author} {\bibinfo {author} {\bibfnamefont {H.}~\bibnamefont
  {Freudenthal}},\ }\href {\doibase 10.1007/978-94-010-2903-2} {\emph {\bibinfo
  {title} {Mathematics as an educational task}}}\ (\bibinfo  {publisher}
  {Springer},\ \bibinfo {year} {1973})\BibitemShut {NoStop}%
\bibitem [{\citenamefont {Treffers}(1987)}]{Adrian1987}%
  \BibitemOpen
  \bibfield  {author} {\bibinfo {author} {\bibfnamefont {A.}~\bibnamefont
  {Treffers}},\ }\href@noop {} {\emph {\bibinfo {title} {Three dimensions: A
  model of goal and theory description in mathematics instruction}}}\ (\bibinfo
   {publisher} {The Wiskobas Project, (D. Reidel Dordrecht},\ \bibinfo {year}
  {1987})\BibitemShut {NoStop}%
\bibitem [{\citenamefont {Brahmia}\ \emph {et~al.}(2015)\citenamefont
  {Brahmia}, \citenamefont {Boudreaux},\ and\ \citenamefont
  {Kanim}}]{brahmia2015}%
  \BibitemOpen
  \bibfield  {author} {\bibinfo {author} {\bibfnamefont {S.}~\bibnamefont
  {Brahmia}}, \bibinfo {author} {\bibfnamefont {A.}~\bibnamefont {Boudreaux}},
  \ and\ \bibinfo {author} {\bibfnamefont {S.~E.}\ \bibnamefont {Kanim}},\
  }\href@noop {} {\bibfield  {journal} {\bibinfo  {journal} {arXiv Preprint.
  arXiv:1602.02033}\ } (\bibinfo {year} {2015})}\BibitemShut {NoStop}%
\bibitem [{\citenamefont {Brahmia}\ \emph {et~al.}(2016)\citenamefont
  {Brahmia}, \citenamefont {Boudreaux},\ and\ \citenamefont
  {Kanim}}]{brahmia2016}%
  \BibitemOpen
  \bibfield  {author} {\bibinfo {author} {\bibfnamefont {S.}~\bibnamefont
  {Brahmia}}, \bibinfo {author} {\bibfnamefont {A.}~\bibnamefont {Boudreaux}},
  \ and\ \bibinfo {author} {\bibfnamefont {S.~E.}\ \bibnamefont {Kanim}},\
  }\href@noop {} {\bibfield  {journal} {\bibinfo  {journal} {arXiv preprint
  arXiv:1601.01235}\ } (\bibinfo {year} {2016})}\BibitemShut {NoStop}%
\bibitem [{\citenamefont {Tuminaro}\ and\ \citenamefont
  {Redish}(2007)}]{Tuminaro2007}%
  \BibitemOpen
  \bibfield  {author} {\bibinfo {author} {\bibfnamefont {J.}~\bibnamefont
  {Tuminaro}}\ and\ \bibinfo {author} {\bibfnamefont {E.~F.}\ \bibnamefont
  {Redish}},\ }\href@noop {} {\bibfield  {journal} {\bibinfo  {journal} {Phys.
  Rev. ST Phys. Educ. Res.}\ }\textbf {\bibinfo {volume} {3}} (\bibinfo {year}
  {2007})}\BibitemShut {NoStop}%
\bibitem [{\citenamefont {Zohrabi~Alaee}\ \emph {et~al.}(2018)\citenamefont
  {Zohrabi~Alaee}, \citenamefont {Sayre},\ and\ \citenamefont
  {Franklin}}]{DZA2018}%
  \BibitemOpen
  \bibfield  {author} {\bibinfo {author} {\bibfnamefont {D.}~\bibnamefont
  {Zohrabi~Alaee}}, \bibinfo {author} {\bibfnamefont {E.}~\bibnamefont
  {Sayre}}, \ and\ \bibinfo {author} {\bibfnamefont {S.}~\bibnamefont
  {Franklin}},\ }in\ \href@noop {} {\emph {\bibinfo {booktitle} {Phys. Educ.
  res. Conf.}}},\ \bibinfo {series and number} {PER Conference}\ (\bibinfo
  {address} {Washington, DC},\ \bibinfo {year} {2018})\BibitemShut {NoStop}%
\bibitem [{\citenamefont {Burton}\ and\ \citenamefont
  {Morgan}(2000)}]{Burton2000}%
  \BibitemOpen
  \bibfield  {author} {\bibinfo {author} {\bibfnamefont {L.}~\bibnamefont
  {Burton}}\ and\ \bibinfo {author} {\bibfnamefont {C.}~\bibnamefont
  {Morgan}},\ }\href@noop {} {\bibfield  {journal} {\bibinfo  {journal} {J.
  Res. Math. Educ.}\ }\textbf {\bibinfo {volume} {31}},\ \bibinfo {pages} {429}
  (\bibinfo {year} {2000})}\BibitemShut {NoStop}%
\bibitem [{\citenamefont {Cope}\ and\ \citenamefont {Kalantzis}(1993)}]{Kress}%
  \BibitemOpen
  \bibfield  {author} {\bibinfo {author} {\bibfnamefont {B.}~\bibnamefont
  {Cope}}\ and\ \bibinfo {author} {\bibfnamefont {M.}~\bibnamefont
  {Kalantzis}},\ }\href@noop {} {\emph {\bibinfo {title} {The powers of
  literacy: A genre approach to teaching writing}}}\ (\bibinfo  {publisher}
  {Falmer},\ \bibinfo {year} {1993})\BibitemShut {NoStop}%
\bibitem [{\citenamefont {Young}(2015)}]{Young2015}%
  \BibitemOpen
  \bibfield  {author} {\bibinfo {author} {\bibfnamefont {H.~D.}\ \bibnamefont
  {Young}},\ }\href@noop {} {\emph {\bibinfo {title} {{University physics with
  modern physics}}}}\ (\bibinfo  {publisher} {Addison--Wesley},\ \bibinfo
  {year} {2015})\BibitemShut {NoStop}%
\bibitem [{\citenamefont {Krane}(1995)}]{Krane1995}%
  \BibitemOpen
  \bibfield  {author} {\bibinfo {author} {\bibfnamefont {S.~K.}\ \bibnamefont
  {Krane}},\ }\href@noop {} {\emph {\bibinfo {title} {{Modern physics}}}}\
  (\bibinfo  {publisher} {Wiley},\ \bibinfo {year} {1995})\BibitemShut
  {NoStop}%
\bibitem [{\citenamefont {Taylor}(2005)}]{Taylor2005}%
  \BibitemOpen
  \bibfield  {author} {\bibinfo {author} {\bibfnamefont {J.~R.}\ \bibnamefont
  {Taylor}},\ }\href@noop {} {\emph {\bibinfo {title} {{Classical
  mechanics}}}}\ (\bibinfo  {publisher} {University Science Books},\ \bibinfo
  {year} {2005})\BibitemShut {NoStop}%
\bibitem [{\citenamefont {Griffiths}(1999)}]{Griffiths1999EM}%
  \BibitemOpen
  \bibfield  {author} {\bibinfo {author} {\bibfnamefont {D.~J.}\ \bibnamefont
  {Griffiths}},\ }\href@noop {} {\emph {\bibinfo {title} {Introduction to
  electrodynamics}}}\ (\bibinfo  {publisher} {Prentice Hall},\ \bibinfo {year}
  {1999})\BibitemShut {NoStop}%
\bibitem [{\citenamefont {Griffiths}(2005)}]{Griffiths2005QM}%
  \BibitemOpen
  \bibfield  {author} {\bibinfo {author} {\bibfnamefont {D.~J.}\ \bibnamefont
  {Griffiths}},\ }\href@noop {} {\emph {\bibinfo {title} {{Introduction to
  quantum mechanics}}}}\ (\bibinfo  {publisher} {Pearson Prentice Hall},\
  \bibinfo {address} {Upper Saddle River, NJ},\ \bibinfo {year}
  {2005})\BibitemShut {NoStop}%
\bibitem [{\citenamefont {Rittle-Johnson}\ and\ \citenamefont
  {Siegler}(1998)}]{rittle1998}%
  \BibitemOpen
  \bibfield  {author} {\bibinfo {author} {\bibfnamefont {B.}~\bibnamefont
  {Rittle-Johnson}}\ and\ \bibinfo {author} {\bibfnamefont {R.~S.}\
  \bibnamefont {Siegler}},\ }\href@noop {} {\emph {\bibinfo {title} {The
  relation between conceptual and procedural knowledge in learning mathematics:
  A review.}}}\ (\bibinfo  {publisher} {The development of mathematical
  skills},\ \bibinfo {year} {1998})\BibitemShut {NoStop}%
\bibitem [{\citenamefont {Russ}\ \emph {et~al.}(2012)\citenamefont {Russ},
  \citenamefont {Lee},\ and\ \citenamefont {Sherin}}]{Russ2012}%
  \BibitemOpen
  \bibfield  {author} {\bibinfo {author} {\bibfnamefont {R.~S.}\ \bibnamefont
  {Russ}}, \bibinfo {author} {\bibfnamefont {V.~R.}\ \bibnamefont {Lee}}, \
  and\ \bibinfo {author} {\bibfnamefont {B.~L.}\ \bibnamefont {Sherin}},\
  }\href {\doibase 10.1002/sce.21014} {\bibfield  {journal} {\bibinfo
  {journal} {Sci. Educ.}\ ,\ \bibinfo {pages} {573}} (\bibinfo {year}
  {2012})}\BibitemShut {NoStop}%
\bibitem [{\citenamefont {Sayre}\ and\ \citenamefont
  {Wittmann}(2008)}]{Sayre2008Coords}%
  \BibitemOpen
  \bibfield  {author} {\bibinfo {author} {\bibfnamefont {E.~C.}\ \bibnamefont
  {Sayre}}\ and\ \bibinfo {author} {\bibfnamefont {M.~C.}\ \bibnamefont
  {Wittmann}},\ }\href@noop {} {\bibfield  {journal} {\bibinfo  {journal}
  {Phys. Rev. ST Phys. Educ. Res.}\ }\textbf {\bibinfo {volume} {4}},\ \bibinfo
  {pages} {20105} (\bibinfo {year} {2008})}\BibitemShut {NoStop}%
\bibitem [{\citenamefont {Sayre}\ and\ \citenamefont
  {Irving}(2015)}]{Sayre2015Brief}%
  \BibitemOpen
  \bibfield  {author} {\bibinfo {author} {\bibfnamefont {E.~C.}\ \bibnamefont
  {Sayre}}\ and\ \bibinfo {author} {\bibfnamefont {P.~W.}\ \bibnamefont
  {Irving}},\ }\href {\doibase 10.1103/PhysRevSTPER.11.020121} {\bibfield
  {journal} {\bibinfo  {journal} {Phys. Rev. ST Phys. Educ. Res.}\ }\textbf
  {\bibinfo {volume} {11}},\ \bibinfo {pages} {020121} (\bibinfo {year}
  {2015})}\BibitemShut {NoStop}%
\end{thebibliography}%

%

\end{document}